\documentstyle[12pt]{article}

\def\np#1#2#3{{\it Nucl. Phys.} {\bf B#1} (#2) #3}
\def\pl#1#2#3{{\it Phys. Lett. }{\bf B#1} (#2) #3}

\def\physrev#1#2#3{{\it Phys. Rev.} {\bf D#1} (#2) #3}

\def\ijmp#1#2#3{{\it Int. J. Mod. Phys.} {\bf #1} (#2) #3}

\def\jhep#1#2#3{{\it JHEP} {\bf #1} (#2) #3}

\def\Ref#1{(\ref{#1})}
\def\plabel#1{\label{#1}}

\def\subsub#1{
\vskip 0.2cm

{\bf #1}
}

\newcommand{\be}{\begin{equation}}
\newcommand{\ee}{\end{equation}}
\newcommand{\beq}{\begin{eqnarray}}
\newcommand{\eeq}{\end{eqnarray}}
\newcommand{\bea}[2]{\be\label{#2}\begin{array}{#1}}
\newcommand{\eea}{\end{array}\ee}

\def\Nb{{\rm \bf N}}

\def\Zb{{\rm \bf Z}}

\def\Re{\,{\rm Re}\, }

\def\rangl{\right\rangle   }
\def\langl{\left\langle  }
\def\({\left(}
\def\){\right)}
\def\[{\left[}
\def\]{\right]}
\def\p{\partial}

\def\11{1\!\! 1}

\def\hf{{1\over 2}}

\def\eps{\varepsilon}

\def\l{\lambda}

\def\s{\sigma}

   \def\CF {{\cal F}}

   \def\CL {{\cal L}}

   \def\CR {{\cal R}}

\def\bz{\bar z}

\def\FSL{\CF}
\def\mul{\mu_{_L}}

\def\ef{e^{-{1\over R}X}}
\def\eg{e^{-{R-1\over R}X}}

\begin{document}

\title{$(m,n)$ ZZ branes and the $c=1$ matrix  model}
\author{Sergei Alexandrov\protect\thanks{e.mail: S.Alexandrov@phys.uu.nl}} 
\date{}

\maketitle

\vspace{-4.5cm}

\hspace{12cm} {ITP-UU-03/52, SPIN-03/33}

\vspace{3cm}

\begin{center}
Institute for Theoretical Physics \& Spinoza Institute, \\
Utrecht University, Postbus 80.195, 3508 TD Utrecht, The Netherlands
\end{center}

\begin{abstract}
We argue that the origin of non-perturbative corrections $e^{-2\pi R n\mu}$ 
in the $c=1$ matrix model is $(1,n)$ D-branes of Zamolodchikovs.
We confirm this identification comparing the flow of these corrections
under the Sine--Liouville perturbation in the two approaches.
\end{abstract}

\section{Introduction}

Recently, a large progress was achieved in understanding of non-perturbative physics
in non-critical string theories 
\cite{MARTINEC,McGreevyKB,KlebanovKM,MTV,KAK,DKKMMS,Gutperle,Teschner,KMS}. 
This progress was related with the discovery of D-branes in Liouville theory. 
However, the remarkable work \cite{ZamolodchikovAH}, where the D-branes were first
introduced, raised also new questions. In that paper a two-parameter family of
consistent boundary conditions describing the D-branes was found,
whereas up to now only one of these solutions, the so called basic $(1,1)$ brane,
played a role in the study of non-perturbative effects. 

In fact, the $(1,1)$ brane is distinguished among the others which are called 
$(m,n)$ branes. There are several reasons for that. In particular, all branes 
except the basic one contain negative dimension operators in their spectrum, 
so that they were thought to be unstable.
Therefore, it is not evident that they play any role at all.
(See, however, \cite{Hos,Pons} where the general case of the $(m,n)$ branes 
is discussed.)

In this letter we address the question: do 
the other $(m,n)$ branes contribute to non-perturbative effects?
We argue that at least the series of $(1,n)$ branes does contribute.
We identify their contributions in the matrix model of the $c=1$ string theory.
Namely, we consider the matrix model of the $c=1$ string with
a Sine-Liouville perturbation and we compare the first two terms 
in the $\lambda$ expansion of the non-perturbative corrections
to the partition function of the perturbed model with the CFT correlators
on the disk with the $(m,n)$ boundary conditions.

The corrections, which we deal with, appear as $e^{-2\pi R n\mu}$ in the formula of
Gross and Klebanov for the unperturbed partition function \cite{GK}.
The first of them with $n=1$ was analyzed in \cite{KAK} and identified with 
the contribution of the $(1,1)$ brane. 
Here we show that the flow of other corrections under the
Sine--Liouville perturbation is exactly the same as one for the $(1,n)$ branes.
Given the results of \cite{KAK}, the calculations are quite simple.
Thus, we conjecture that at least in the $c=1$ string theory
the $(1,n)$ branes are associated with the higher non-perturbative corrections
to the partition function.

The plan of the paper is the following. In the next section we review the 
relevant properties of the $(m,n)$ ZZ branes. In section 3 we review the results 
of \cite{KAK} on non-perturbative corrections in the perturbed $c=1$ theory.
Finally, in section 4 we present our evidences in favour of the identification of the 
$(1,n)$ branes with the non-perturbative corrections coming from 
the higher exponential terms in the formula of
Gross and Klebanov. 
In Appendix A the behaviour of these corrections near the $c=0$
critical point is studied.

\section{D-branes in Liouville theory}

Liouville theory appears when one considers (a matter coupled to)
two-dimensional gravity in the conformal gauge. 
It is defined by the following action:
\be
S_L= \int {d^2\s\over 4\pi} \left( (\p\phi )^2 + Q\hat R\phi
+\mul e^{2b \phi} \right).
\plabel{LIOU}
\ee
The central charge of this CFT is given by
\be
c_L=1+6Q^2
\plabel{cliouv}
\ee
and the parameter $b$ is related to Q via the relation
\be
Q=b+{1\over b}.
\plabel{bQ}
\ee
In general, $b$ and $Q$ are determined by the requirement that the 
total central charge
of matter and the Liouville field is equal to $26$. 
In the case when matter is represented by a 
minimal $(p,q)$ model with the central charge 
$c_{p,q}=1-6{(p-q)^2\over pq}$, the relation \Ref{cliouv} implies that
\be
b=\sqrt{p/q}.
\plabel{bminmod}
\ee
The coupling to the $c=1$ matter corresponds to the limit $b\to 1$.

An important class of conformal primaries in Liouville 
theory corresponds to the operators
\be
V_\alpha(\phi)=e^{2\alpha\phi}
\plabel{opalph}
\ee
whose scaling dimension is given by $\Delta_\alpha=
\bar\Delta_\alpha=\alpha(Q-\alpha)$.
The Liouville interaction in \Ref{LIOU} is $\delta\CL=\mul V_b$.

It is known that D-branes in Liouville theory
should be localized in the strong coupling region $\phi\to\infty$.
The reason is that in this region the energy of the brane, 
which is proportional to the inverse string coupling $1/g_s\sim e^{-Q\phi}$, 
takes its minimum. 
Thus, at the classical level a configuration describing an open string attached 
to such a D-brane 
should be represented by the Liouville field on a disk, 
which goes to infinity approaching the boundary of the disk.
The classical Liouville action \Ref{LIOU} does have such a solution. 
It describes a surface of constant negative curvature. 
Realized as the unit disk $|z|<1$ on the complex plane, it has the following metric 
\be
ds^2=e^{\phi(z)}|dz|^2, \qquad 
e^{\phi(z)}={1\over \pi\mul b^2}{1\over (1-z\bz)^2}.
\plabel{lobg}
\ee

The quantization of this geometry was constructed in the work
\cite{ZamolodchikovAH}. It was shown that there is a two-parameter family
of consistent quantizations which are in one-to-one correspondence
with the degenerate representations of the Virasoro algebra.
Thus, one can talk about $(m,n)$ D-branes ($m,n\in \Nb$) of Liouville theory.
We list the main properties of these branes 
which were established in \cite{ZamolodchikovAH}.

\subsub{Quasiclassical behaviour.} 
Only the set of $(1,n)$ branes has a smooth behaviour in the limit $b\to 0$ and
can be interpreted as quantization of the classical geometry \Ref{lobg}.

\subsub{Perturbative expansion.} 
Only the $(1,1)$ solution is consistent with the loop perturbation theory.

\subsub{Spectrum of boundary operators.}
Each $(m,n)$ boundary condition is associated with a
boundary operator represented by the corresponding
degenerate field of Liouville theory, so that the spectrum of representations
of open strings between two D-branes is defined by the fusion rules of these
fields. In particular, the (1,1) boundary condition
contains only the identity operator, whereas all other branes include operators
of negative dimensions in their spectrum.

\subsub{One-point correlation functions.}
The {\it unnormalized} one-point correlators of the primary operators 
$V_{\alpha}$ \Ref{opalph}
on the disk with the $(m,n)$ boundary conditions are given by the boundary 
wave function $\Psi_{m,n}(P)$ with momentum $iP=\alpha-Q/2$. 
The explicit form of these functions will not be important for us.
But we need the fact that their dependence of $m$ and $n$ is contained
in the simple factor so that
\be
\Psi_{m,n}(P)=\Psi_{1,1}(P)
{\sinh(2\pi mP/b)\sinh(2\pi nbP)\over \sinh(2\pi P/b)\sinh(2\pi bP)}.
\plabel{onep}
\ee

\section{Non-perturbative effects in the compactified $c=1$ string theory}

In this section we review the results of \cite{KAK} on
the leading non-perturbative effects
in two dimensional string theory with a winding perturbation. In the world sheet
description, the Lagrangian of this system is 
\be
\CL={1\over 4\pi}\left[(\partial x)^2 +(\partial\phi)^2
+2\hat \CR\phi+\mul \phi e^{2\phi}+\lambda e^{(2-R)\phi}
\cos [R(x_L-x_R)]\right],
\plabel{conepert}
\ee
where the field $x$ is compactified on a circle of
radius $R$, $x\simeq x+2\pi R$. 
The Sine--Liouville interaction described by the last
term gives rise to vortices on the string world sheet. The
perturbation is relevant for $R<2$, and we will restrict to the case
$R\in (1,2) $ in the discussion below.

The analysis of \cite{KAK} was based on the matrix model of the CFT \Ref{conepert}.
This model was constructed in \cite{KazakovPM}
where it was shown that the Legendre transform $\FSL$ 
of the string partition sum satisfies the Toda differential equation
\be
{1\over 4}\lambda^{-1}\p_{\lambda}\lambda\p_{\lambda} \FSL(\mu, \lambda)+
\exp\left[-4\sin^2\(\hf{\p\over \p\mu}\) \FSL(\mu,\lambda)\right]=1
\plabel{todafd}
\ee
with initial condition provided by the partition function of the 
unperturbed $c=1$ string theory on a circle \cite{GK}
\beq
& \FSL(\mu,0)  = 
{R\over 4} \Re \int_{\Lambda^{-1}}^\infty {ds\over s}
{ e^{-i\mu s}\over \sinh{s\over 2}\sinh{s\over 2R}} &
\nonumber \\
& = -  {R\over2}\mu^2 \log {\mu\over\Lambda} -
{1\over24}\big(R + {1\over R} \big)\log {\mu\over\Lambda} +
R\sum_{h=2}^\infty \mu^{2-2h} c_h(R)+O(e^{-2\pi\mu})+O(e^{-2\pi R \mu}).&
\plabel{FrenO}
\eeq
Here the genus $h$ term $c_h(R)$ is a known polynomial in ${1/ R}$. 

The genus expansion of the function $\FSL(\mu,\lambda)$ can be found
solving the Toda equation. In the following we
will need only the genus-0 solution \cite{KazakovPM}.
To present it, it is convenient to introduce the scaling parameters
\be
y=\mu\xi, \qquad \xi=(\lambda\sqrt{R-1})^{-{2\over 2-R}}~.
\plabel{scp}
\ee
In terms of these variables the second derivative of the
partition sum on the sphere, $\FSL_0(\mu,\l)$, reads
\be
\p^2_{\mu}\FSL_0= R\log\xi+X(y),
\plabel{Fzero}
\ee
where the function $X(y)$ is determined by the following algebraic equation
\be
 y=\ef-\eg.  
\plabel{wf}
\ee

At $\lambda=0$, the non-perturbative corrections 
to the perturbative expansion of the partition function follow from
the formula \Ref{FrenO} of Gross and Klebanov.
They are associated with the poles of the integrand
in that equation, which occur at $s=2\pi ik$ and
$s=2\pi R ik$, $k\in \Zb$. Correspondingly, there are two series 
of the non-perturbative corrections, $\exp(-2\pi k\mu)$ and
$\exp(-2\pi R k\mu)$. The leading ones with $k=1$ are indicated 
on the second line of \Ref{FrenO}.

The dependence of these non-perturbative corrections 
on $\lambda$ in the presence of the Sine--Liouville
interaction was analyzed in \cite{KAK}.
It was shown that the first series of 
the corrections is not modified by the perturbation
and a non-trivial flow is related to the second series.

In \cite{KAK} the following problem was solved.
Let a non-perturbative correction to the partition function has the
following exponential form
\be
\eps(\mu,y)\sim e^{-\mu f(y)}~. 
\plabel{anz}
\ee
Then from the fact that the full partition function should be still a solution of the
Toda equation \Ref{todafd} (in the spherical approximation), the 
function $f(y)$ was found. 
The solution is written in the following parametric form
\beq
&
f(y)=2\phi(y)+ {2(2-R)\over y\sqrt{R-1}}e^{-\hf X(y)}\sin \phi(y), &
\plabel{fphi}
\\
&
{ \cos \({1\over R}  \phi(y)-\psi\)
\over \cos \( {R-1\over R} \phi(y)+\psi\) }
=-{1\over\sqrt{R-1}}\, \exp\( {2-R\over 2R}X(y)\). &
\plabel{solz}
\eeq
Note that the function $\phi(y)$ is directly related to the derivative of $f(y)$:
\be
\phi(y)=\hf \p_y(yf(y)).
\plabel{fph}
\ee
The constant $\psi$ is determined by the initial condition.
If this condition is fixed by the unperturbed theory, $\psi$ should be found
from the given asymptotics at $y\to\infty$.

In \cite{KAK} the flow with $\lambda$ of the first correction $\exp(-2\pi R\mu)$
was investigated. It corresponds to the initial condition 
$\lim\limits_{y\to\infty}f(y)=2\pi R$ which fixes the constant in \Ref{solz} as follows
\be
\psi={\pi \over 2}.
\plabel{inpsi}
\ee
Thus, the equation \Ref{solz} takes the form
\be
{ \sin \({1\over R} \phi\)
\over \sin \( {R-1\over R} \phi\) }= {1\over\sqrt{R-1}}\, e^{{2-R\over 2R}X(y)}.
\plabel{zzz}
\ee

Also several first terms of the expansion of $f(y)$  
for small $\lambda$, or large $y$, were found
\be
f(y)= 2\pi R + {4\sin(\pi R)}\, \mu^{-{2-R\over 2}}\, \lambda +
R\sin(2\pi R)\, \mu^{-(2-R)}\,\lambda^2+O(\lambda^3).
\plabel{corrf}
\ee
%

\section{$(1,n)$ branes in the matrix model}

In \cite{KAK} it was demonstrated that the first two terms of the expansion \Ref{corrf}
give rise to non-perturbative 
effects which are reproduced from D-brane calculations in the CFT framework
based on the (1,1) ZZ brane. This confirmed the proposal that the basic $(1,1)$
brane is responsible for the leading non-perturbative corrections.
But the result \Ref{FrenO} implies that there is a series of such corrections, whereas
only the first one was analyzed. Here we are going to identify the origin
of all other corrections. 

In terms of the function $f(y)$ introduced in the previous section,
they are described by the initial conditions
\be
\lim\limits_{y\to\infty}f_k(y)=2\pi R k.
\plabel{ink}
\ee
The flow with the Sine--Liouville coupling is determined by equations
\Ref{fphi} and \Ref{solz}. It is clear that all initial conditions \Ref{ink}
correspond to the same constant $\psi$ given in \Ref{inpsi}. As a result,
all functions $f_k(y)$ are defined by the same set of equations \Ref{fphi}
and \Ref{zzz}.

It is trivial to generalize the expansion \Ref{corrf} to the case of arbitrary parameter $k$.
The result reads
\be
f_k(y)
= 2\pi R k + {4\sin(\pi R k)}\, \mu^{-{2-R\over 2}}\, \lambda +
R\sin(2\pi R k)\, \mu^{-(2-R)}\,\lambda^2+O(\lambda^3).
\plabel{corrfk}
\ee
Thus, in each order in $\lambda$ the $k$-dependence of the strength of the non-perturbative
corrections is quite simple. In particular, we have
\be
\left.{f_k\over f_1}\right|_{\lambda=0}=k, 
\qquad
\left.{\p_{\lambda}f_k\over \p_{\lambda} f_1}\right|_{\lambda=0}
={\sin (\pi R k)\over \sin(\pi R)}.
\plabel{relf}
\ee

On the other hand, these non-perturbative corrections should be related to
one-point correlators of the cosmological and Sine--Liouville operators, respectively,
on the disk with some boundary conditions. As we mentioned, for $k=1$ 
it was shown that these are the $(1,1)$ boundary conditions.
Here we argue that for general $k$ one should choose the $(1,k)$ boundary conditions.

We can give two evidences in favour of this conjecture. 
Namely, assuming this identification,
one can reproduce the two relations \Ref{relf}. Indeed, it implies that
\beq
\left.{f_k\over f_1}\right|_{\lambda=0}&=&
{\langl V_b\rangl_{1,k}\over \langl V_b\rangl_{1,1}}=
{\Psi_{1,k}(i(Q/2-b))\over \Psi_{1,1}(i(Q/2-b))}, 
\plabel{relPsi}
\\
\left.{\p_{\lambda}f_k\over \p_{\lambda} f_1}\right|_{\lambda=0}&=&
{\langl V_{b-R/2}\rangl_{1,k}\over \langl V_{b-R/2}\rangl_{1,1}}=
{\Psi_{1,k}(i(Q/2-b+R/2))\over \Psi_{1,1}(i(Q/2-b+R/2))},
\plabel{relPsil}
\eeq
where in the right hand side one should take the limit $b\to 1$.
Taking into account \Ref{onep} and \Ref{bQ}, it is easy to show that
one obtains the matrix model result \Ref{relf}.

It is clear that this result can be reproduced also from $(k,1)$ ZZ branes.
However, we refer to the absence of a sensible quasiclassical limit for these
branes to argue that the $(1,k)$ branes are more preferable.

\section{ Discussion }

Several remarks concerning our conjecture are in order.

First, if one looks at the degenerate fields of Liouville theory 
\be 
\Phi_{m,n}=\exp\(\( (1-m)/b+(1-n)b\)\phi \)
\ee
and at their dimensions
\be
\Delta_{m,n}=Q^2/4-(m/b+nb)^2/4,
\ee
one observes that
in the limit $b\to 1$ all degenerate fields with fixed $m+n$ are indistinguishable.
Therefore, one can wonder how we were able to resolve this indeterminacy
for the ZZ branes.
It is not completely clear for us why, but the formula \Ref{onep} shows
that even in this limit all $(m,n)$ branes are different
since we obtain either the product $\sinh(2\pi mP)\sinh(2\pi nP)$ or simply $mn$ 
for the case $P=0$. The only remaining symmetry is the possibility to exchange 
$m$ and $n$. Thus, in the $c=1$ theory it is possible that 
the branes $(m,n)$ and $(n,m)$ are indeed the same. If, nevertheless, they are different,
we can apply the argument given in the end of the previous section
to choose between $(1,n)$ and $(n,1)$ branes.

One can ask also what happens in the $c<1$ case.
In this case the KP equations of the matrix model should produce the similar
series of non-perturbative corrections to the partition function as the one appearing
in the equation \Ref{FrenO}. Namely, they will have the form 
$e^{-kf_{p,q}}$, $k\in \Nb$.\footnote{In fact, 
for each minimal $(p,q)$ model there is a finite
set of such corrections which are in one-to-one correspondence with the Kac table.
The reason for the appearance of the additional parameters, let us call them $(r,s)$,
is that the minimal CFT model possesses exactly the same number 
of possible Dirichlet boundary conditions \cite{CARDY}. For $q=p+1$, all constants
$f_{p,q}(r,s)$ were found in \cite{EynardSG}.}
However, the factor $k$ is not reproduced by any $(m,n)$ brane.
Indeed, taking into account the relation \Ref{bminmod}, one obtains that for the $(p,q)$
minimal model 
\be 
{\langl V_b\rangl_{m,n}\over \langl V_b\rangl_{1,1}}=
{\Psi_{m,n}(i(Q/2-b))\over \Psi_{1,1}(i(Q/2-b))}=
(-1)^{m+n}{\sin(\pi mq/p)\sin(\pi np/q)\over \sin(\pi q/p)\sin(\pi p/q)}. 
\plabel{relPsimin}
\ee
No choice of $m$ and $n$ makes this expression equal to $k$.
Thus, the origin of these non-perturbative corrections is not the same as in the $c=1$ string
theory.

The latter statement is confirmed by the analysis of the non-perturbative
corrections in the Sine--Liouville theory
near the $c=0$ critical point, which is presented in Appendix A.
On could expect that the $k$th correction in the $c=1$ string theory flows to the $k$th
correction in the $c=0$ theory.
However,  it turns out that only the leading correction possesses the correct $c=0$ behaviour
and reproduces the leading correction to the pure gravity partition function.
In contrast, all corrections with $k>1$ die off in this limit.

In fact, the corrections, which are just $k$th power of the leading correction, can be 
interpreted as contributions of $k$ instantons described by the $(1,1)$ 
branes.\footnote{This fact was pointed out to us by I. Kostov.}
This interpretation works well both for the $c<1$ and the unperturbed $c=1$ cases.
However, when the Sine--Liouville interaction is taken into account, 
this picture breaks down.
The reason is that it implies the existence of the exponential corrections, which
are powers of the leading one, independently of the presence
of a perturbation because effects of interaction between instantons can appear only
in the next orders in genus expansion. But, as we saw above, this is not true for the $c=1$
theory perturbed by windings where the flow with $\lambda$ leads to a more
complicated $k$-dependence (see eq. \Ref{relf}). Thus, it is not clear whether
even for $c<1$ the higher corrections can be explained by
the many instanton interpretation. Another possibility would be 
that in the $c=1$ theory the $(1,n)$ branes are bound states of n $(1,1)$ branes
which are stabilized by the perturbation.

It would be quite interesting to do 
additional checks of the conjecture we proposed in this letter.
One of them is the comparison of the two-point correlator of
the Sine--Liouville operator on the disk with 
the $(1,n)$ boundary conditions to the third term in the $\lambda$
expansion in \Ref{corrfk}. In fact, we need only to know the $(m,n)$ dependence
of this correlator to check the conjecture.

\bigskip
\noindent{\bf Acknowledgements:}
The author is grateful to  V. Kazakov, I. Kostov, B. Ponsot and J. Teschner
for valuable discussions and especially to D. Kutasov for discussion of the $c=0$
critical limit. 
The author thanks the organizers of the workshop "Integrable Models and Applications:
From Strings to Condensed Matter" in Florence 
(Euclid Network HPRN-CT-2002-00325)
for hospitality during the beginning of this work.

\appendix

\section{$c=0$ critical behaviour}

It is well known that decreasing the cosmological coupling $\mu$
in the Sine--Liouville theory \Ref{conepert}, one  
approaches a critical point where the system, which had the central charge 1,
behaves as pure gravity with the vanishing central charge \cite{HsuCM}.
In \cite{KAK} it was shown that the same is true for the non-perturbative effects.
Namely, in the limit 
\be
y\longrightarrow
y_c=-(2-R)(R-1)^{R-1\over 2-R}
\plabel{yyccrr}
\ee
the function $f_1(y)$ describing the leading non-perturbative correction 
to the partition function of the Sine--Liouville theory reproduces
the leading non-perturbative correction to the partition function of the pure gravity.

In this appendix we analyze what happens for the higher corrections $f_k(y)$.
They are defined by the same pair of equations \Ref{fphi} and \Ref{zzz} and differ only
by the following condition
\be
\lim\limits_{y\to\infty}\phi_k(y)=\pi R k.
\plabel{phik}
\ee
However, in the limit \Ref{yyccrr} one finds a crucial difference between
the case $k=1$ and all other cases. Indeed, the equation \Ref{wf} implies that the critical value 
of the function $X$ is given by 
\be
e^{-{2-R\over 2R}X_c}=\sqrt{R-1}.
\plabel{xxccrr}
\ee
Thus, the critical values of $\phi_k$ are to be found from the following equation
\be
{ \sin \({1\over R} \phi_k^{(c)}\)
\over \sin \( {R-1\over R} \phi_k^{(c)}\) }= {1\over R-1}.
\plabel{zcr}
\ee
For $k=1$ the solution of this equation is evident:
$\phi_1^{(c)}= 0$. But for all $k>1$ (and generic values of $R$), the critical value is 
non-vanishing and belongs to the interval $\(\pi R (k-1), \pi R (k+1)\)$.

This fact has important consequences. Whereas for $k=1$ the expansion near the critical point
gives \cite{KAK}
\be
\plabel{crphi}
\phi_1(y)=\sqrt{3}(X_c-X)^{1/2}+O\((X_c-X)^{3/2}\),
\ee
for $k>1$ one finds\footnote{There is no misprint here. The second order term vanishes indeed.} 
\be
\phi_k(y)=\phi_k^{(c)}-{2-R\over 2} {\sin \({1\over R} \phi_k^{(c)}\) \over
\cos \({1\over R} \phi_k^{(c)}\) -\cos  \( {R-1\over R} \phi_k^{(c)}\) }\,
(X_c-X)+O\((X_c-X)^{3}\).
\plabel{crphik}
\ee
Substitution of these expansions into the general solution \Ref{fphi} 
together with the critical behaviour of the free energy 
\be
X_c-X={\sqrt{2}R\over\sqrt{R-1}}\(y_c-y\over y_c\)^{1/2} +O\(y_c-y\) ,
\plabel{yXrel}
\ee
gives two different results\footnote{In fact, the high order terms are easier calculated 
taking into account the relation \Ref{fph}.} 
\beq
f_1(y)&\approx& - {8\sqrt{3}\over 5 }\({2 R^2\over R-1}\)^{1/4}
\({y_c-y \over y_c}\)^{5/4}+\cdots,
\plabel{focrit}  \\
f_{k}(y)&\approx&2(\phi_k^{(c)}-\sin \phi_k^{(c)}) 
 -2\sin \phi_k^{(c)}\({y_c-y \over y_c}\) 
\nonumber \\
&+&{2\sqrt{2} R(2-R)\over 3\sqrt{R-1}} {\sin \({1\over R} \phi_k^{(c)}\) \over
\cos \({1\over R} \phi_k^{(c)}\) -\cos  \( {R-1\over R} \phi_k^{(c)}\) }\,
\({y_c-y \over y_c}\)^{3/2}+\cdots, \quad k>1.
\plabel{fkcrit}
\eeq

Thus, all higher corrections possess a critical behaviour which is different from the usual
$c=0$ behaviour given by the first correction. 
What is the final point of their RG flow?
In the critical limit for $k>1$ the leading two terms of the non-perturbative corrections are
\be
\eps_k\sim \exp\(-\mu f_{k}(y)\)\sim
\exp \(-{2y_c \over \xi }(\phi_k^{(c)}-\sin \phi_k^{(c)}) 
 -{2\phi_k^{(c)} \over \xi }(y-y_c) \).
\plabel{fykcrit}
\ee
The $c=0$ string coupling is related to the critical parameter as follows
\be
g_s \sim {\xi\over (y_c-y)^{5/4}}\, .
\plabel{gscr}
\ee
Thus, to keep it fixed or, moreover, small, one should take also $\xi\to 0$.
Then the first term goes to $+\infty$ (since $y_c<0$) and the correction 
apparently diverges.  However, this term does not depend on any parameters 
of the $c=0$ model and can be absorbed into the normalization coefficient
which is, in any case, arbitrary.
The next term does depend on $g_s$ and leads to the following behaviour
\be
\eps_k \sim \exp\(-c_k g_s^{-4/5}\), \qquad c_k \sim {\phi_k^{(c)}\over \xi^{1/5}}.
\plabel{kcor}
\ee
As a result, in the limit we are interested in, $c_k\to \infty$. 
Therefore, in contrast to the leading correction, which reproduces
the non-perturbative correction of the $c=0$ theory, all higher
corrections disappear near the critical point.


\end{document}